\documentclass[12pt, letterpaper]{article}

\usepackage{amsmath, amssymb}
\usepackage{cite}
\usepackage{fancyhdr}
\usepackage[top=1in, bottom=1.5in, left=1in, right=1in]{geometry}
\usepackage[dvipdfm, dvips]{graphicx}
\usepackage{hyperref}

\numberwithin{equation}{section}

\begin{document}


\setcounter{page}{0}
\date{}

\lhead{}\chead{}\rhead{\footnotesize{UCI-HEP-TR-2011-01\\UCSD-PTH-11-04}}\lfoot{}\cfoot{}\rfoot{}

\title{\textbf{Collider Constraints on\vspace{0.2cm}\\
Dipole-Interacting Dark Matter \vspace{0.4cm}}}

\author{Jean-Fran\c{c}ois Fortin$^1$ and Tim M.P. Tait$^2$\vspace{0.7cm}\\
{\normalsize{$^1$Department of Physics, University of California, San Diego}}\\
{\normalsize{La Jolla, CA 92093-0319, USA}}\vspace{0.2cm}\\
{\normalsize{$^2$Department of Physics and Astronomy, University of California, Irvine}}\\
{\normalsize{Irvine, CA 92697-4575, USA}}}

\maketitle
\thispagestyle{fancy}

\begin{abstract}
\normalsize
\noindent
Dark matter which interacts through a magnetic or electric dipole moment is an interesting possibility which may help to resolve the discrepancy between the DAMA annual modulation signal and the null results of other searches.  In this article we examine relic density and collider constraints on such dark matter, and find that for couplings needed to explain DAMA, the thermal relic density is generically in the right ballpark to account for cosmological measurements.  Collider constraints are relevant for light WIMPs, but less constraining that direct searches for masses above about 10 GeV.
\end{abstract}

\newpage
\tableofcontents

\vspace{1cm}

\section{Introduction}\label{sec:intro}

There is overwhelming evidence that the Universe is filled with nonbaryonic dark matter (DM).  However, despite the wealth of cosmological evidence for its existence, its particle physics properties remain a complete mystery.  Among the many candidates proposed, weakly interacting massive particles (WIMPs) are among the best motivated, due to the possibility of understanding their abundance as thermal relics and potential connection to electroweak symmetry-breaking.  The hunt is on to discover dark matter through direct observation of its scattering with nuclei, indirectly through its annihilation into high energy particles, and through its production at colliders.

This is a particularly exciting time for direct detection experiments.  Detectors are achieving unprecedented sensitivity, and will improve their discovery reach by orders of magnitude.  A positive signal at such an experiment would be our first hint that dark matter has interactions beyond gravitational, and would confirm the WIMP picture of dark matter and open the door to improving our quantitative understanding of the early Universe through its relic density.  In fact, the experimental situation is somewhat confusing.  DAMA/LIBRA \cite{Bernabei:2010mq} and CoGeNT 
\cite{Aalseth:2010vx} both report signals which have defied explanation in terms of conventional backgrounds to WIMP searches, but XENON \cite{Aprile:2010um}, and CDMS \cite{Ahmed:2009zw} seem to exclude some or all of the parameter space necessary to explain them.

One of the key properties which distinguish different visions for the nature of dark matter are its spin and the nature of its interactions with Standard Model (SM) particles.  The most common assumption is that heavy particles (including perhaps the SM $Z$ and Higgs bosons) mediate the interactions, resulting in interactions which are essentially contact interactions at the length scales probed by the direct searches.  However, it remains possible that dark matter interacts with ordinary matter by exchanging photons.  To remain dark, it must be electrically neutral, so the coupling to photons must be described by a higher multipole interaction.  In particular, a Dirac fermion WIMP could couple to photons through an electric or magnetic dipole moment, and this could provide the dominant mechanism for scattering off heavy nuclei either elastically
\cite{Raby:1987ga,Bagnasco:1993st,Sigurdson:2004zp,Masso:2009mu,Barger:2010gv,Fitzpatrick:2010br,Banks:2010eh} or inelastically \cite{Chang:2010en}.

In Ref.~\cite{Banks:2010eh} it was found that dipole-moment interacting dark matter (DMDM) with mass around $10$ GeV and magnetic gyromagnetic ratio around $0.02$ (i.e. with magnetic dipole moment around $10^{-17}$ $e$-cm) is (marginally) consistent with results from direct detection experiments including CDMS-II \cite{Ahmed:2009zw}, XENON100 \cite{Aprile:2010um}, DAMA \cite{Bernabei:2010mq}, and CoGeNT \cite{Aalseth:2010vx}.  
As part of this work, we update these bounds to include the recent CDMS low threshold analysis \cite{Ahmed:2010wy} and find that it excludes the CoGeNT and DAMA regions for elastic scattering at the $90\%$ CL.  Nonetheless, the electromagentic dipole moment is an interesting portal for a WIMP to interact with the Standard Model, and it is an important part of the over-all picture of dark matter interactions to understand the bounds from existing experiments as well as future prospects for discovery.

A final important question is whether or not DMDM can realize the WIMP(less) miracle \cite{Feng:2008ya}.  Early proposals for DMDM had large dipole moments by virtue of the WIMP itself being composite \cite{Banks:2005hc}.  The dark matter relic abundance was generated non-thermally through strong dynamics, and ultimately determined by a primordial asymmetry between WIMPs and anti-WIMPS.  This is an attractive picture, but the magnetic Land\'{e} factor consistent with the direct detection experiments is small enough to suggest that the dark sector is weakly coupled.  This opens the possibility that the dark matter may be a more standard thermal relic, and it would be interesting to see how the parameter space consistent with a thermal relic is related to that relevant for direct detection experiments.

In this article, we build upon Refs \cite{Bagnasco:1993st,Sigurdson:2004zp,Masso:2009mu,Barger:2010gv,Banks:2010eh} to explore DMDM from the points of view of the thermal relic density and in light of constraints from both LEP and hadron colliders \cite{Goodman:2010yf,Goodman:2010ku,Bai:2010hh,Cao:2009uw}.  It is organized as follows: Section~\ref{sec:DMDM} briefly reviews the theory of DMDM.  Section~\ref{sec:Abundance} computes the thermal relic density of DMDM and identifies the parameter space consistent with a thermal relic.  Section \ref{sec:Collider} discusses collider bounds on DMDM from LEP and the Tevatron as well as the reach of the LHC.  Finally Section \ref{sec:conc} summarizes the DMDM bounds from direct detection experiments \cite{Banks:2010eh}, the DMDM thermal relic density constraint for weakly-coupled dark sectors and the DMDM bounds from colliders.


\section{Dark Matter with Dipole Moments}\label{sec:DMDM}

We work in an effective theory framework, assuming that the only relevant degree of freedom of the dark sector at the energies of interest is the WIMP itself.  The WIMP is a Dirac fermion DM $\psi$ of mass $m_{\rm DM}$ which interacts with the SM through electromagnetic dipole moments \cite{Bagnasco:1993st,Sigurdson:2004zp,Masso:2009mu,Barger:2010gv,Banks:2010eh},
\begin{equation}\label{eqn:effectiveLagrangian}
\mathcal{L}_{\rm DM}=
\bar{\psi}(i\gamma^\mu\partial_\mu-m_{\rm DM})\psi
+\frac{g_Me}{8m_{\rm DM}}\bar{\psi}\sigma^{\mu\nu}\psi F_{\mu\nu}
+\frac{g_Ee}{8m_{\rm DM}}\bar{\psi}\sigma^{\mu\nu}\psi\widetilde{F}_{\mu\nu}
\end{equation}
where $g_M$ and $g_E$ are the DM magnetic and electric gyromagnetic ratios, and $F_{\mu \nu}$ and $\widetilde{F}_{\mu\nu}$ are the electromagnetic (dual) field strengths.  Along with the Higgs portal operator $|H|^2 \bar{\psi} \psi$ (considered elsewhere \cite{Kanemura:2011nm}), the magnetic and electric dipole operators are the gauge-invariant dimension $5$ operators coupling the DM to the SM and are thus among the most relevant operators mediating SM-DM interaction.

The electric dipole moment operator violates the discrete symmetries P and T.  In models where these symmetries are spontaneously broken, the electric dipole operator is effectively of dimension $5+n$ operator where $n\geq1$.  The magnetic and electric Land\'{e} factors are related to the DM scale $\Lambda_{\rm DM}$ and CP-violating scale $\Lambda_{\rm CP}>\Lambda_{\rm DM}$ through $g_M\sim4m_{\rm DM}/\Lambda_{\rm DM}$ and $g_E\sim4m_{\rm DM}/\Lambda_{\rm DM}(\Lambda_{\rm DM}/\Lambda_{\rm CP})^n$ respectively.

In terms of the parameters of $g_M$ and $g_E$, the DMDM has a magnetic dipole moment $\mu_{\rm DM}=g_Me/4m_{\rm DM}$ and an electric dipole moment 
$d_{\rm DM}=g_Ee/4m_{\rm DM}$.  Naive dimensional analysis suggests that fermionic DM Land\'{e} factors $g_M$ and $(\Lambda_{\rm CP}/\Lambda_{\rm DM})^ng_E$ are order one in theories with a strongly-coupled dark sector, and loop-suppressed when the dark sector is weakly-coupled.


\section{Relic Abundance}\label{sec:Abundance}

In this section we examine the thermal relic density of DMDM in theories with a weakly-coupled dark sectors.  We perform the thermal relic density computation under the assumption that there is no primordial asymmetry between $\psi$ and $\bar{\psi}$.  At freeze-out, the WIMP is non-relativistic for the masses of interest, and its relic density is determined from its annihilation cross section into Standard Model particles as usual \cite{Kolb:1988aj}.

The annihilation cross sections for DMDM to SM particles at lowest order in the relative velocity $v_{\rm rel}$ and neglecting the fermion masses are given by\footnote{Note that the work of \cite{Masso:2009mu} does not take into account annihilation into photons.}
\begin{equation}\label{eqn:DM-SM}
\begin{array}{rclcrcl}
\sigma_{\bar{\psi}\psi\rightarrow\bar{f}f}^{\rm MDM}v_{\rm rel} &=& 
\frac{\pi(g_MQ_f\alpha)^2}{4m_{\rm DM}^2} 
&\hspace{2cm}& \sigma_{\bar{\psi}\psi\rightarrow\bar{f}f}^{\rm EDM}v_{\rm rel} &=& 
\frac{\pi(g_EQ_f\alpha)^2}{48m_{\rm DM}^2}v_{\rm rel}^2\vspace{0.2cm}\\
\sigma_{\bar{\psi}\psi\rightarrow\gamma\gamma}^{\rm MDM}v_{\rm rel} &=& 
\frac{\pi(g_M^2\alpha)^2}{64m_{\rm DM}^2} 
&\hspace{2cm}& \sigma_{\bar{\psi}\psi\rightarrow\gamma\gamma}^{\rm EDM}v_{\rm rel} &=& 
\frac{\pi(g_E^2\alpha)^2}{64m_{\rm DM}^2}
\end{array}
\end{equation}
where $Q_f$ is the fermion charge.  In the limit of zero SM fermion mass, magnetic-dipole interacting dark matter has predominantly $s$-wave annihilation into SM fermions, while the leading electric-dipole interaction is $p$-wave.  Annihilation into photons is $s$-wave for both.  We neglect annihilation into a pair of $W$ bosons and/or top quarks, which will correct these expressions for $m_{\rm DM} \gtrsim 80$~GeV, but as we will see, the possibility of a thermal relic for this range of masses is already excluded by direct detection experiments.

The thermally averaged annihilation cross section $\sigma_A$ can be written,
\begin{equation}\label{eqn:sigmaA}
\langle\sigma_Av_{\rm rel}\rangle=
\sum_{m_f<m_{\rm DM}}\langle\sigma_{\bar{\psi}\psi\rightarrow\bar{f}f}v_{\rm rel}\rangle
+\langle\sigma_{\bar{\psi}\psi\rightarrow\gamma\gamma}v_{\rm rel}\rangle
\equiv \sigma_0 
\left[1+b\left(\frac{T}{m_{\rm DM}}\right)
\right]
\end{equation}
where $\sigma_0$ and $b$ are extracted in terms of $g_M$ or $g_E$ and $m_{\rm DM}$ from Eq.~(\ref{eqn:DM-SM}) (and $b=0$ for magnetic interactions).  

To good approximation, the freeze-out temperature $T_f$ is given by \cite{Kolb:1988aj}
\begin{eqnarray}\label{eqn:Tf}
x_f=\frac{m_{\rm DM}}{T_f} &=& \ln[0.038~(g/g_*^{1/2})m_{\rm Pl}m_{\rm DM}\sigma_0]\nonumber\\
 && -\frac{1}{2} \ln \{ \ln[0.038~(g/g_*^{1/2})m_{\rm Pl}m_{\rm DM}\sigma_0]\}\nonumber\\[0.2cm]
 && +\ln(1+b\{\ln[0.038(g/g_*^{1/2})m_{\rm Pl}m_{\rm DM}\sigma_0]\}^{-1}),
\end{eqnarray}
where $m_{\rm Pl}$ is the Planck mass, $g=2$ is the number of degrees of freedom of the DM and $g_*$ is the effective number of relativistic degrees of freedom at the time of decoupling.  The DMDM relic density is then \cite{Kolb:1988aj,Servant:2002aq}
\begin{equation}\label{eqn:Omegahsq}
\Omega_{\rm DM}h^2=
\frac{1.07\times 10^9 ~{\rm GeV}^{-1}}{m_{\rm Pl}} ~
\frac{x_f}{\sigma_0 \left[1+ 3 b x_f^{-1} \right](g_{*S}/g_*^{1/2})}
\end{equation}
where $g_{*S}$ is the effective number of degrees of freedom relevant for the entropy at the time of decoupling.  At the time of freeze-out, all particle species have a common temperature and  $g_{*S}=g_*$.  To saturate the observed DM relic density, DMDM which is a thermal relic should satisfy $\Omega_{\rm DM}h^2=0.1120\pm0.0056$ \cite{Komatsu:2010fb}.  The values of $g_M$ ($g_E$) for which the correct thermal relic density is obtained for a give WIMP mass are plotted in Figures~\ref{fig:90pclm} and~\ref{fig:90pcle}, respectively.  



\section{Collider Constraints}\label{sec:Collider}

DMDM can also be produced at colliders, which may provide complimentary information to the direct detection experiments \cite{Banks:2010eh}.  The analysis is somewhat similar to studies which probe contact interactions between WIMPs and quarks and gluons \cite{Goodman:2010yf,Goodman:2010ku,Bai:2010hh} or leptons \cite{Fox:2011fx},\footnote{For example, \cite{Fox:2011fx} investigates WIMP interactions through heavy mediators with leptons, whereas we consider WIMPs interacting through photons.} but with the added feature that the mediator particle is now the massless photon.  We examine the implications of collider searches for photon plus missing energy and jet plus missing energy final states on DMDM.

Any theory with non-renormalizable interactions must be understood as an effective theory, breaking down at some finite energy scale.  Firm predictions can only be made at energies well below this scale.  In the case of DMDM, theories with a weakly coupled dark sector such that $g_M, g_E \ll 1$ have robust predictions for WIMP production at the LHC.  For a strongly coupled dark sector such as a composite WIMP, the momentum transfer will be of order the compositeness scale, which could invalidate bounds (and also could potentially lead to a variety of other interesting signals).

\subsection{LEP II Constraints}\label{subsec:Lepton}

In lepton colliders,  production of $\psi \bar{\psi}$ (missing transverse energy in the detector) plus a photon can lead to an observable signal,
\begin{equation*}
e^-+e^+\rightarrow\bar{\psi}+\psi+\gamma.
\end{equation*}
The photon can be produced either as initial state radiation from the incoming leptons, or directly from the DMDM final state through is dipole moment.  Initial state radiation enjoys a large collinear singularity which causes the cross section to grow as $\ln E_\gamma / m_e$ when the photon is approximately collinear with the incoming $e^\pm$ beams.  Thus, it completely dominates the final state radiation (which we neglect).  The center-of-mass production cross section for this process is given in the massless electron limit by
\begin{eqnarray*}
\frac{d\sigma_\gamma}{dx_{E_T^\gamma}d\eta_\gamma} &=& \frac{g^2\alpha^3}{12s}\sqrt{1-\frac{4x_{m_{\rm DM}}^2}{1-2x_{E_T^\gamma}\cosh(\eta_\gamma)}}\\
 && \times\frac{[1-2x_{E_T^\gamma}\cosh(\eta_\gamma)+2(1+3\sigma)x_{m_{\rm DM}}^2][1-2x_{E_T^\gamma}\cosh(\eta_\gamma)+x_{E_T^\gamma}^2\cosh(2\eta_\gamma)]}{x_{E_T^\gamma}x_{m_{\rm DM}}^2[1-2x_{E_T^\gamma}\cosh(\eta_\gamma)]}
\end{eqnarray*}
where $\sqrt{s}$ is the center-of-mass energy, $E_T^\gamma=x_{E_T^\gamma}\sqrt{s}$ is the photon transverse energy, $\cos(\theta_\gamma)=\tanh(\eta_\gamma)$ is the photon polar angle, and $m_{\rm DM}=x_{m_{\rm DM}}\sqrt{s}$.  For DMDM with magnetic dipole moment $g=g_M$ and $\sigma=1$ while for DMDM with electric dipole moment $g=g_E$ and $\sigma=-1$.  Note that the full cross section with massive electrons is used in generating the plots.

A search for this signal was performed using the L3 detector based on several different LEP II collider energies \cite{Achard:2003tx}.  The most significant SM background is production of a $Z$-boson (decaying to neutrinos, and this producing missing transverse energy in the final state)
together with initial state radiation.  Events with missed particles or particles with mismeasured energies also contribute a ``fake" background.  Of these, Bhabha events where both charged leptons are lost turn out to be the most important \cite{Achard:2003tx}.


L3 considered two event topologies containing one final state photon \cite{Achard:2003tx}:
\begin{itemize}
\item High energy single photon:
\begin{itemize}
\item[-]$E_T^\gamma>0.02\sqrt{s}$;
\item[-]$14^\circ<\theta_\gamma<166^\circ$;
\item[-]No other photon with $E_\gamma>1$ GeV.
\end{itemize}
\item Low energy single photon:
\begin{itemize}
\item[-]$0.008\sqrt{s}<E_T^\gamma<0.02\sqrt{s}$;
\item[-]$43^\circ<\theta_\gamma<137^\circ$;
\item[-]No other photon with $E_\gamma>1$ GeV.
\end{itemize}
\end{itemize}
The data set is divided in $8$ subsets with specific luminosities and center-of-mass energies $\sqrt{s}$ between $189$ and $209$ GeV.  The signal efficiencies vary between $69.8\%$ and $73.7\%$ across these data sets.  We choose a flat efficiency of $71\%$ close to the median value for all energies.  We simulate events and use the LEP bound to derive limits on the cross section.  To be conservative, we derive $90\%$ CL bounds from the single event topology and data subset that leads to the best constraints, and do not attempt to combine the various datasets.

Our results are summarized on Figures~\ref{fig:90pclm} and \ref{fig:90pcle}, from which we see that bounds from LEP II are effective up to about the kinematic limit of $m_{\rm DM} \lesssim 100$~GeV.  For both types of interaction, the LEP limits do better than those from direct detection for masses less than about 5 GeV, but in neither case is LEP able to probe couplings small enough to explain the thermal relic density.


\subsection{Hadron Collider Constraints}\label{subsec:Hadron}

In hadron colliders, fermionic DMDM can be searched for in photon plus missing transverse energy and jet plus missing transverse energy final states.  We focus here on the jet plus missing transverse energy final state
\begin{equation*}
p^++p^\pm\rightarrow\bar{\psi}+\psi+{\rm jet}.
\end{equation*}
Photon plus missing transverse energy final states are interesting but the rate is suppressed by the electromagnetic coupling $\alpha$ compared to the jet plus missing transverse energy process which is proportional to the strong coupling $\alpha_S$.  In estimating our signal rates, we use MSTW parton distribution functions (PDFs) \cite{Martin:2009iq}.  The center-of-mass production cross section for this process, in which the final jet can originate from a quark(anti-quark) or a gluon at the parton level, is given in the massless parton limit by
\begin{eqnarray*}
\frac{d\sigma_{q-{\rm jet}}}{dx_{E_T^{\rm jet}}d\eta_{\rm jet}} &=& C_qQ^2\frac{g^2\alpha^2\alpha_S}{24s}\sqrt{1-\frac{4x_{m_{\rm DM}}^2}{1-2x_{E_T^{\rm jet}}\cosh(\eta_{\rm jet})}}\\
 && \times\frac{[1-2x_{E_T^{\rm jet}}\cosh(\eta_{\rm jet})+2(1+3\sigma)x_{m_{\rm DM}}^2]}{x_{m_{\rm DM}}^2[1-2x_{E_T^{\rm jet}}\cosh(\eta_{\rm jet})]\cosh(\eta_{\rm jet})[1+\tanh(\eta_{\rm jet})]}\\
 && \times\{1-2x_{E_T^{\rm jet}}\cosh(\eta_{\rm jet})[1-\tanh(\eta_{\rm jet})]\\
 && \hspace{1cm}+x_{E_T^{\rm jet}}^2\cosh^2(\eta_{\rm jet})[5-2\tanh(\eta_{\rm jet})+\tanh^2(\eta_{\rm jet})]\}\\
\frac{d\sigma_{g-{\rm jet}}}{dx_{E_T^{\rm jet}}d\eta_{\rm jet}} &=& C_gQ^2\frac{\alpha_S}{\alpha}\frac{d\sigma_\gamma}{dx_{E_T^{\rm jet}}d\eta_{\rm jet}}
\end{eqnarray*}
where again $\sqrt{s}$ is the center-of-mass energy, $E_T^{\rm jet}=x_{E_T^{\rm jet}}\sqrt{s}$ is the jet transverse energy, $\cos(\theta_{\rm jet})=\tanh(\eta_{\rm jet})$ is the jet polar angle, $m_{\rm DM}=x_{m_{\rm DM}}\sqrt{s}$, $Q$ is the initial parton electric charge and $C_q,C_g$ are the appropriate color factors.  As usual $g=g_M$ and $\sigma=1$ for DMDM with magnetic dipole moment while $g=g_E$ and $\sigma=-1$ for DMDM with electric dipole moment.

The most significant SM background comes from $Z$-bosons decaying to neutrinos with associated jet production.  Background also includes events with missed particles or particles with mismeasured energies.  Of these, events leading to $W$-boson plus jets where the charged lepton from the $W$-boson decay is missed are the most important \cite{Aaltonen:2008hh,CDF}.

\subsubsection*{Tevatron $90\%$ Confidence Limits}

Both D0 and CDF searched for the jet plus missing transverse energy final state, inspired by models with large extra dimensions.  We compare with the CDF \cite{Aaltonen:2008hh,CDF} search, which is based on a significantly larger integrated luminosity and thus provides the strongest limits.
The CDF event selection included \cite{CDF}:
\begin{itemize}
\item[-]$E_T^{\rm jet}>80$ GeV;
\item[-]$|\eta_{\rm jet}|<1.1$;
\item[-]$\not{\!\!E_T}>80$ GeV;
\item[-]At most one extra jet with $E_T<30$ GeV;
\item[-]Any additional jet with $E_T>20$ GeV vetoed.
\end{itemize}
The SM predicted $8663$ events in the $1$ fb$^{-1}$ of data at $\sqrt{s}=1.96$ TeV studied while only $8449$ events were observed by CDF.  The signal event efficiency was found in \cite{Goodman:2010ku} to be roughly $40\%$, rather independent of the WIMP mass.  From Figures~\ref{fig:90pclm} and \ref{fig:90pcle}, we see that bounds from the Tevatron are weaker than those from LEP II for masses smaller than about 100 GeV, where direct detection experiments are providing the best limits, several orders of magnitude better than those from colliders.

\subsubsection*{LHC $5\sigma$ Reach}

Predictions for the jet plus missing transverse energy final state was studied at the LHC in \cite{Vacavant:2001sd} (also in the context of large extra dimensions).  The study \cite{Vacavant:2001sd} selected events with:
\begin{itemize}
\item[-]$E_T^{\rm jet}>100$ GeV;
\item[-]$|\eta_{\rm jet}|<3.2$;
\item[-]$\not{\!\!E_T}>500$ GeV.
\end{itemize}
At center-of-mass energy $\sqrt{s}=14$ TeV, $100$ fb$^{-1}$ of data lead to an expected number of background events of approximatively $B=2\cdot10^4$.  To compare with these bounds we follow the results of \cite{Goodman:2010ku} which found an efficiency of $80\%$.  We define a $5\sigma$ discovery region in which DMDM can be discovered in end-stage LHC running.


\section{Discussion and Conclusion}\label{sec:conc}

\begin{figure}[t]
\begin{center}
\includegraphics[scale=0.60]{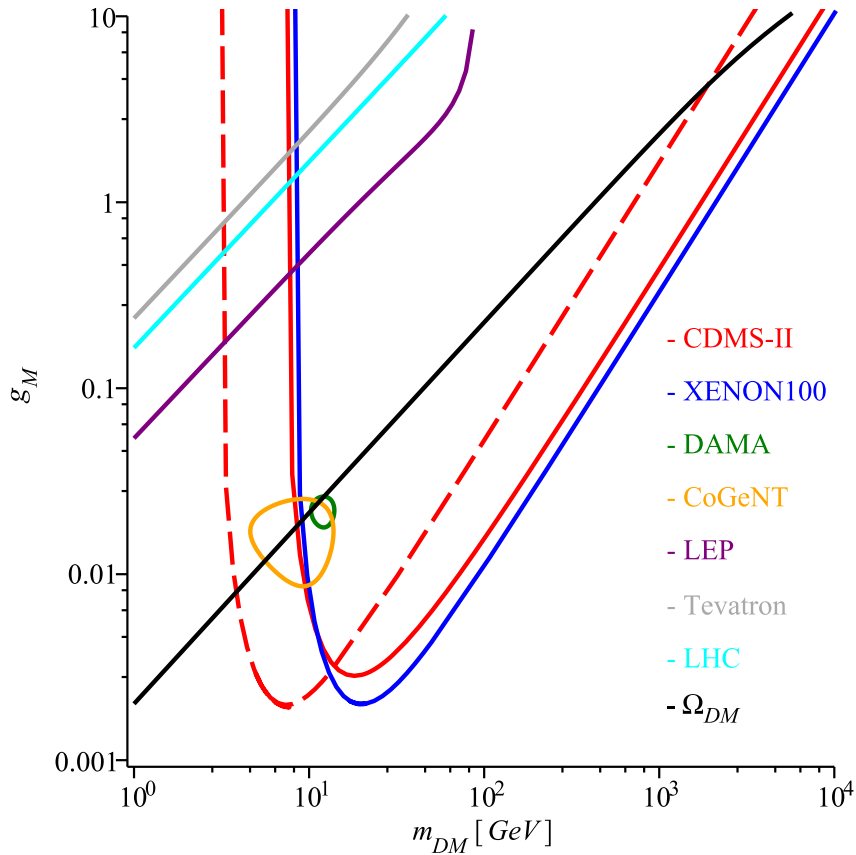}
\end{center}
\caption{$90\%$ confidence level (CDMS-II, XENON100, DAMA, CoGeNT, LEP, Tevatron) and $5\sigma$ reach (LHC) plots for direct detection and collider experiments for DMDM with magnetic dipole moment.  The dash line corresponds to the $90\%$ confidence level plot for the low threshold CDMS analysis.}
\label{fig:90pclm}
\end{figure}

\begin{figure}[t]
\begin{center}
\includegraphics[scale=0.60]{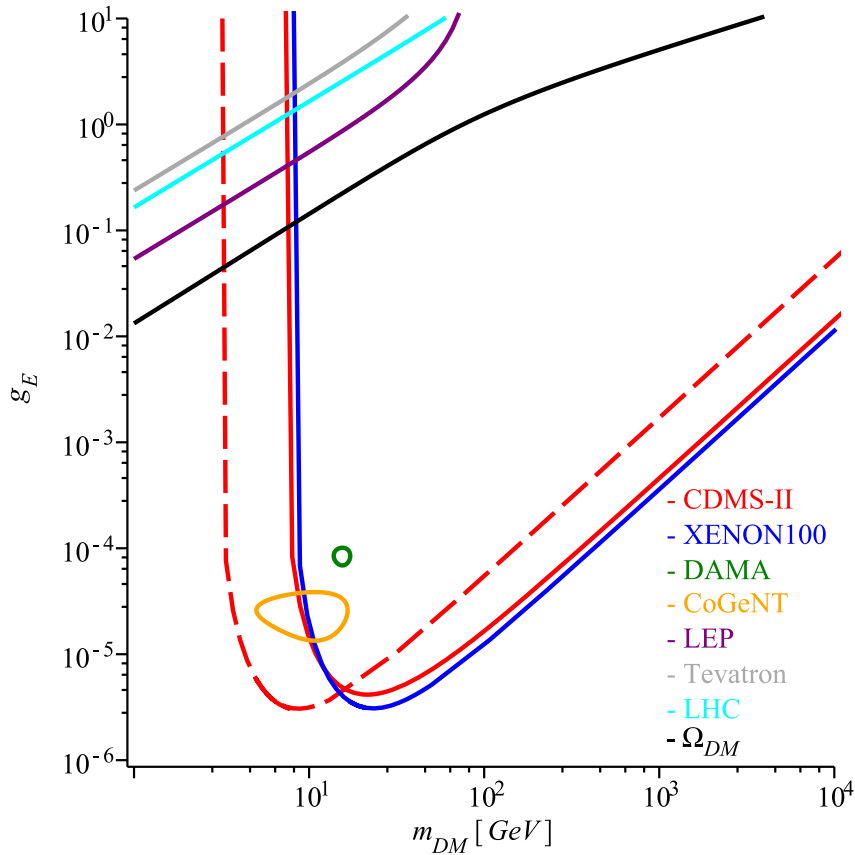}
\end{center}
\caption{$90\%$ confidence level (CDMS-II, XENON100, DAMA, CoGeNT, LEP, Tevatron) and $5\sigma$ reach (LHC) plots for direct detection and collider experiments for DMDM with electric dipole moment.  The dash line corresponds to the $90\%$ confidence level plot for the low threshold CDMS analysis.}
\label{fig:90pcle}
\end{figure}

We have considered a dark matter particle which interacts with the Standard Model through an electric or magnetic dipole moment.  This is an interesting portal connecting the dark sector to ordinary matter, mediated by the massless photon.  As such, it is also one of the most challenging cases for colliders, because its rate drops rapidly with the mass of the WIMP.  

We have considered updated direct detection bounds from CDMS, the thermal relic density, and collider constraints from LEP II and the Tevatron.  We have also considered the long-term LHC prospects for a discovery in the channel of $\psi \bar{\psi} +$~jet.  Our results are summarized in Figures~\ref{fig:90pclm} and \ref{fig:90pcle}.

For masses between a few to 100 GeV, direct detection constraints are already quite strong \cite{Banks:2010eh}, somewhat stronger on the electric dipole moment than on the magnetic dipole moment.  This is a consequence of the low energy enhancement of the electric dipole moment-mediated scattering (see, for example, \cite{Banks:2010eh}), and results in direct detection experiments discriminating between the two.  Conversely, at colliders the dark matter is produced at high energy, and the rates are comparable for both interaction types.  A thermal relic is consistent with all constraints only for masses of order a few GeV, and in the case of a magnetic dipole interaction, would be consistent with the regions favored by the DAMA and CoGeNT signals, had that same region not been (marginally) excluded by the low threshold CDMS analysis.  It is interesting that light dark matter with a dipole moment induced at the loop level seems to pass all of the usual criteria to be a good WIMP\footnote{We note in passing that it is difficult to explain the PAMELA \cite{:2008zzr} positron excess.  To explain PAMELA the total annihilation cross section must satisfy $\langle\sigma_Av_{\rm rel}\rangle\gtrsim10^{-23}$ cm$^3$/s.  For $m_{\rm DM}\lesssim50$ GeV this translates into $g_M\gtrsim0.05m_{\rm DM}/$GeV and $g_E\gtrsim0.76\sqrt{m_{\rm DM}/{\rm GeV}}$.  Since LEP requires $g_{M,E}\lesssim0.05m_{\rm DM}/$GeV for $m_{\rm DM}\lesssim50$ GeV, the former is tightly constrained by LEP while the latter is excluded by LEP.}.  At larger masses, a non-thermal production mechanism such as could occur in theories with a strongly-coupled dark sector \cite{Banks:2005hc} must be invoked.

Collider bounds are typically somewhat weaker than the direct detection bounds, which are also stronger than typical indirect detection bounds from annihilation into $\gamma \gamma$ \cite{Sigurdson:2004zp,Goodman:2010qn}.  Although, due to a cleaner environment, the LEP bounds are stronger by an order of magnitude compared to the Tevatron bounds, both bounds are barely consistent with a weakly-coupled theory.  The LHC discovery prospects in the jets + missing energy channel are not very promising, being already excluded by LEP II at low masses, and direct detection experiments at large masses.  Of course, with its large center-of-mass energy, one could hope that the LHC could also directly produce even heavier states in the dark sector.  Their signatures are model-dependent, but there is much potential for discovery of states with higher production rates and/or more striking experimental signals with lower backgrounds.

Dark matter coupled through the photon portal is an interesting and motivated vision which leads to a somewhat different perspective from the standard WIMP.  Understanding its allowed parameter space is an important step in its search.

\subsection*{Acknowledgments}

We would like to thank Kunal Kumar, Patipan Uttayarat, and especially Wouter Waalewijn for valuable discussions.  The research of JFF was supported in part by DOE grant DOE-FG03-97ER40546.  The research of TT was supported in part by NSF grant PHY-0970171 and in addition he gratefully acknowledges the hospitality of the SLAC theory group, where part of this work was completed.



\end{document}